\documentclass[%
 reprint,
 amsmath,amssymb,
 aps,
pra,
]{revtex4-1}

\usepackage{graphicx}
\usepackage{dcolumn}
\usepackage{subfigure}
\usepackage{bm}
\usepackage{braket}             
\usepackage{amsmath,amsfonts,amssymb}
\usepackage{xcolor}
\usepackage[english]{babel}
\usepackage{lipsum}
\usepackage{comment}
\usepackage{relsize}   

\DeclareMathOperator{\Tr}{Tr}

\begin{document}

\title{The value of the early-time Lieb-Robinson correlation function for qubit arrays}  

\author{Brendan J. Mahoney}
\author{Craig S. Lent}
\affiliation{ 
Department of Electrical Engineering\\
University of Notre Dame\\
Notre Dame, IN 46556, USA
}%

\date{\today}

\begin{abstract}
The Lieb-Robinson correlation function captures propagation  of  quantum correlations in many-body system. We calculate the value of the leading order of the correlation function, not its bound, for a system of interacting qubits at early times. The general analytical result is compared to numerical calculations and is applied to regular qubit lattices in one, two, and three dimensions. We find an exact expression value of the Lieb-Robinson velocity for near-neighbor coupled systems and see the approximately exponential leading edge of correlations emerge for large arrays. 
\end{abstract}


\maketitle






\section{Introduction}

Entanglement in many-body quantum systems and the spread of quantum correlations has been of considerable interest for both fundamental reasons and for the possible applications in quantum computing. Focus on quantum information naturally employs entropic measures derived from the von Neumann entropy of a state. Multiple measures have been proposed to capture the quantum entanglement of spatially separated systems and this remains an area of great activity \cite{Plenio2007,HorodeckiRMP2009,Vedral2008}.
Information so conceived is a property of the {\em state} of the system, characterized by either a state vector for pure states or a density operator for pure or mixed states. 

Another avenue, not  tied to the state of the system, considers the commutation relations between local operators that act on spatially separated subsystems. Preeminent here is investigation of the  Lieb-Robinson operator \cite{LiebRobinson1972}, which quantifies the quantum correlation between a Heisenberg operator $\hat A_j(t)$ which acts on  subsystem $j$ of the composite many-body system, and an operator $\hat B_k$ which acts on a distinct subsystem $k$. The operator is given by  
\begin{equation}
\hat{C}_{A_j,B_k}(t) =  [\hat A_j(t),\hat B_k] 
\label{eq:GenLiebRobOp}
\end{equation}
\noindent The Lieb-Robinson correlation function is given by the norm of the  operator. 
\begin{equation}
{C}_{A_j,B_k}(t) = \left\| \hat{C}_{A_j,B_k}(t)  \right\| 
\label{eq:GenLiebRobF}
\end{equation}
At $t=0$, the operators $\hat A$ and $\hat B$, having different support, commute and the correlation function is zero. We can think of the  operator $\hat A_j(t)$ as spreading out into the system and as its support comes to include that of $\hat B$ the correlation increases. This captures the quantum correlation, between the two systems which is the source of entanglement. The Lieb-Robinson correlation function depends only on the choice of the operators and the Hamiltonian of the system, independent of any initial state.

Lieb and Robinson established, for quite general conditions, that the correlation so defined propagates outward with a finite speed and with an exponentially bounded leading edge in space:
\begin{equation}
    C_{AB}(t) \leq c \,e^{-a(d(j,k) -v_{\text{\tiny LR}} t)}
    \label{eq:LRbound}
\end{equation}
where $d(j,k)$ is an appropriate distance measure between the two subsystems and 
$v_{\text{\tiny LR}}$ is the Lieb-Robinson velocity. 

Because of the importance of this result, there has been a substantial effort to refine the bound in particular circumstances \cite{Hazzard2020, Monroe2010, Yao2020,Kalamara2010,Sims2006,Hastings2004} and to understand the connection between entropic measures and correlation measures \cite{Verstraete2006,Kastner2017}.  Spin models and equivalent qubit arrays have played a prominent role in studying bounds for specific systems \cite{Palencia2021} and can also be  explored experimentally  \cite{King2021}. 

Following the work of Luitz et al. \cite{Luitz2017,Luitz2019,Luitz2019b,LuitzPRR2020}, we focus here on a specific class of Hamiltonians  and compute the correlation function itself, rather than a bound on the correlation function. This comes at the cost of generality, of course, but yields insight into the way correlations propagate.  In Section \ref{sec:LinearArray} we consider  a linear array of near-neighbor coupled qubits---the 1D tranverse  field Ising model \cite{Palencia2021}, and calculate the Lieb-Robinson correlation function numerically. This is only practical for small systems. We then derive an analytical expression for the correlation function which we can compare with the numerical results. The analysis is extended to an arbitrary array of qubits in Section \ref{sec:ArbArray}, and then applied to regular square lattices in two and three dimensions in Section \ref{sec:RegularLattices}. 

For the purposes of stating a bound, the norm in Eq. (\ref{eq:LRbound}) need not be specified, but to calculate the value of the correlation function a particular norm must be chosen.  Here  we  use the normalized Frobenius norm 
\begin{equation}
 \left\| \hat Q  \right\| = \sqrt{\frac{\Tr\left(  \hat Q^\dagger  \hat Q \right)}{\mathcal{N}}},
\label{eq:FrobNorm}
\end{equation}
where $\mathcal{N}$ is the dimension of the Hilbert space. The normalization ensures that the value is independent of the size of the space. 





\vfill
\section{Lieb Robinson correlation function for a regular linear array \label{sec:LinearArray}}
\subsection{Direct numerical calculation}
We consider first a linear array of $N_q$ qubits with an energy $\Delta$ coupling adjacent z-components, as described by the following Hamiltonian:
\begin{equation}
\hat{H} =  - \gamma \sum\limits_k^{{N_q}} {\hat{\sigma}^{(k)}_x - \frac{\Delta }{2}} \sum\limits_k^{{N_q} - 1} {\hat{\sigma}^{(k)}_z} \,\hat{\sigma}^{(k + 1)}_z.
\label{eq:HIsing}
\end{equation}
Here $\hat{\sigma}^{(k)}_\alpha$ is the Pauli spin operator with $\alpha=(x, y, z)$ operating on the $k^{th}$ qubit. Each operator is understood to be embedded in the full direct-product Hilbert space of the array with dimension $\mathcal{N}=2^{N_q}$.
This system is the 1D transverse field Ising model. The first term represent the internal dynamics which allow each qubit to flip its state. The energy $\gamma$ sets the characteristic time for the dynamics 
\begin{equation}
\tau\equiv \pi \hbar/\gamma.
\label{eq:tauDef}
\end{equation}

\noindent We have the usual commutation relations for Pauli operators.
\begin{eqnarray}
\left[\hat\sigma_x^{(j)}, \hat\sigma_y^{(k)}\right] &=& 2\,i\,\hat\sigma_z^{(k)}
   \delta_{j,k} \nonumber\\
\left[\hat\sigma_y^{(j)}, \hat\sigma_z^{(k)}\right] &=& 2\,i\,\hat\sigma_x^{(k)} \delta_{j,k}
\nonumber\\
\left[\hat\sigma_z^{(j)}, \hat\sigma_x^{(k)}\right] &=& 2\,i\,\hat\sigma_y^{(k)} \delta_{j,k}
\end{eqnarray}

We consider the Lieb-Robinson operator between the $z$ components of the first qubit and qubit $k$.
\begin{equation}
\hat{C}_k(t) =  {[\hat{\sigma}_{1}^{z}(t),\hat{\sigma}_{k}^{z}(0)]} 
\label{eq:1DLiebRobOp}
\end{equation}
The Lieb-Robinson correlation function is given by the norm of the  operator. 

\begin{equation}
C_k(t) = \left\| \hat{C}_k(t)  \right\|
\label{eq:1DLiebRobCorF}
\end{equation}

There is no loss of generality here by choosing the first qubit as the reference  because the coupling is the same between any adjacent qubits. The correlation function  $C_k$  captures the quantum correlation between two qubits that are $(k-1)$ apart.

The time-development of the  Heisenberg operator $\hat{\sigma} _{1}^{(z)}$ is given by
\begin{equation}
\hat{\sigma}_z^{(1)}(t) = e^{i\frac{\hat{H}}{\hbar} t} \hat{\sigma}_z^{(1)}
e^{-i\frac{\hat{H}}{\hbar} t}
\label{eq:VonNeumannEquation}
\end{equation}
The correlation function can then be written

\begin{equation}
    C_k(t) = \left\|  \left[ 
            e^{i\frac{\hat{H}}{\hbar} t} \hat{\sigma}_z^{(1)}
            e^{-i\frac{\hat{H}}{\hbar} t},
               \hat{\sigma}_z^{(k)}(0)\right] \right\|.
               \label{eq:LRcorrExact}
\end{equation}

\begin{figure}[tb]
\centering
\includegraphics[width=8.6cm]{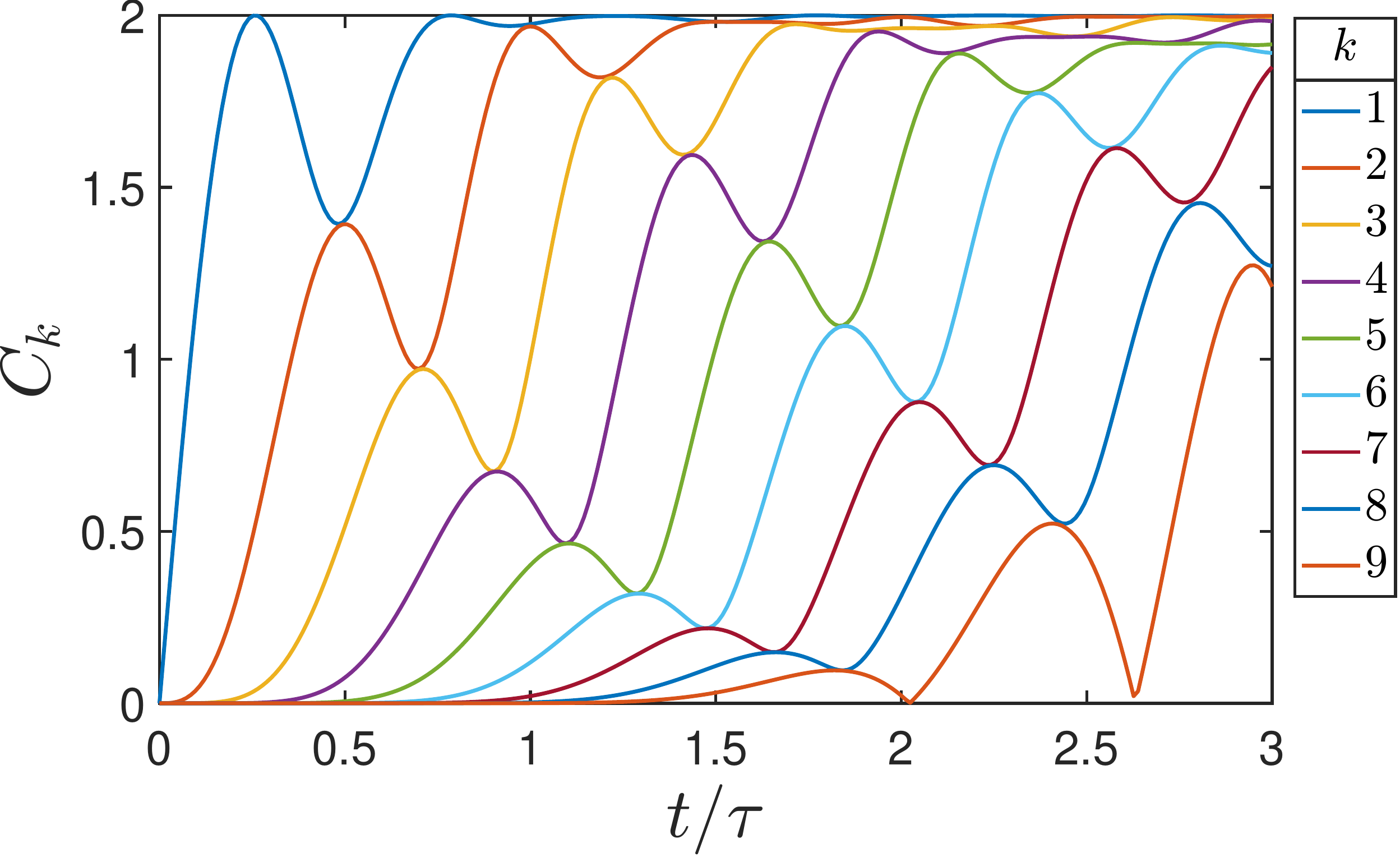}
\caption{The Lieb-Robinson correlation function between the reference qubit 1 and qubit $k$ for a linear array of 9 qubits with near-neighbor interactions. The calculation is done by  numerical evaluation of $C_q$ using  Eq. (\ref{eq:LRcorrExact}) and the Hamiltonian of Eq. (\ref{eq:HIsing}) with $\Delta/\gamma=1.$
}
\label{fig:LRCorrZoomOut}
\end{figure}

Figure \ref{fig:LRCorrZoomOut} shows the  Lieb Robinson correlation function  for a line of nine qubits by  a direct numerical evaluation of Eq. (\ref{eq:LRcorrExact}) for the Hamiltonian in Eq. (\ref{eq:HIsing}). 

\begin{figure}[tb]
\centering
\includegraphics[width=8.6cm]{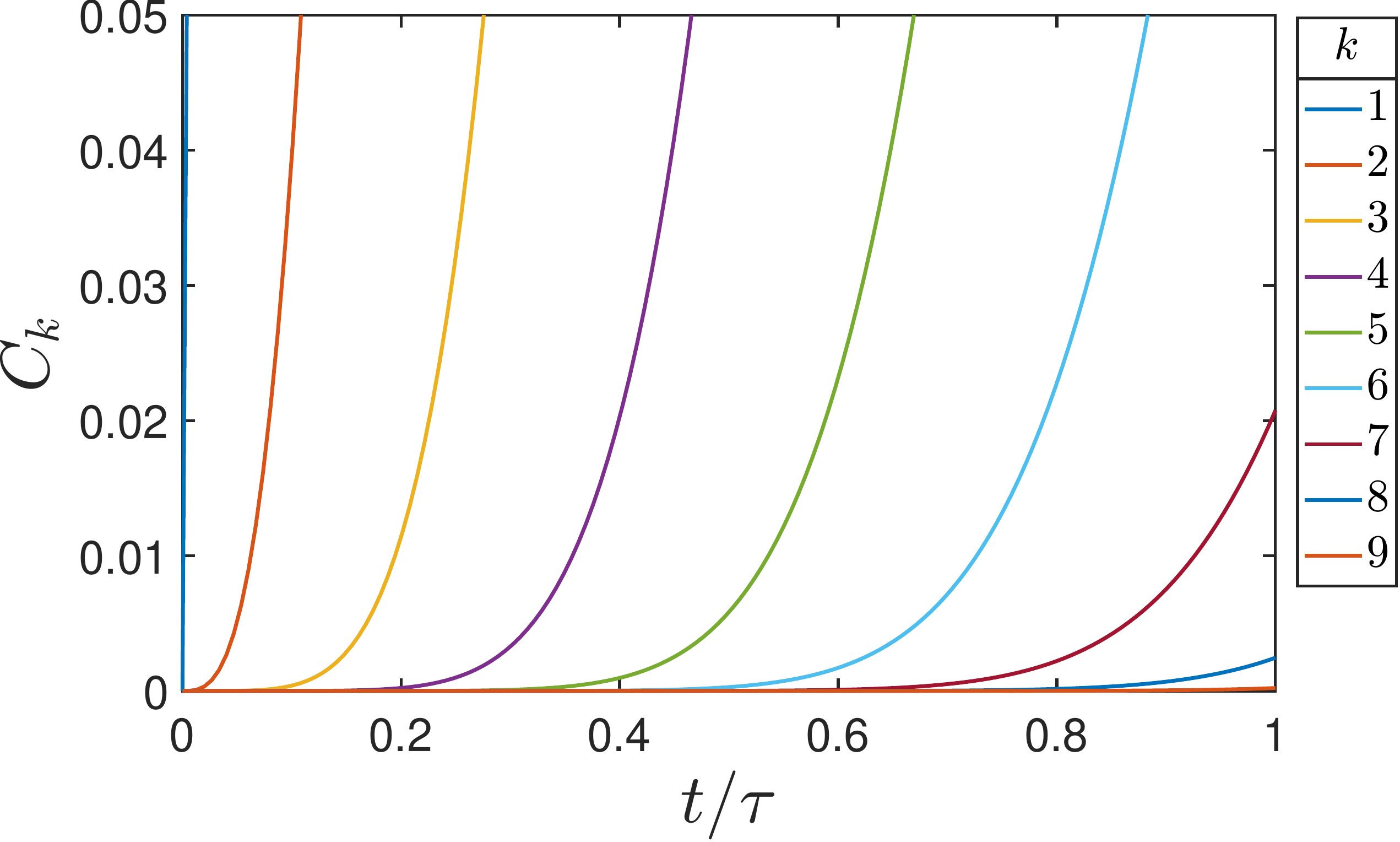}
\caption{Detail of the early-time behavior Lieb-Robinson correlation function as the quantum correlations move down the line of nine qubits. The calculation and parameters are as in Figure (\ref{fig:LRCorrZoomOut}).
}
\label{fig:LRCorrZoomIn}
\end{figure}

We are interested in the early-time behavior as correlations initially spread down the line. By ``early-time'' behavior we mean times such that the correlation with the reference qubit 1 is  small. Our focus is on the leading order term in time as the correlations between the first site and the $k^{th}$ site initially grow.  Figure \ref{fig:LRCorrZoomIn} shows a view of the correlations zoomed in to the lower left-hand corner of Fig. \ref{fig:LRCorrZoomOut}.

\begin{figure}[tb]
\centering
\includegraphics[width=8.6cm]{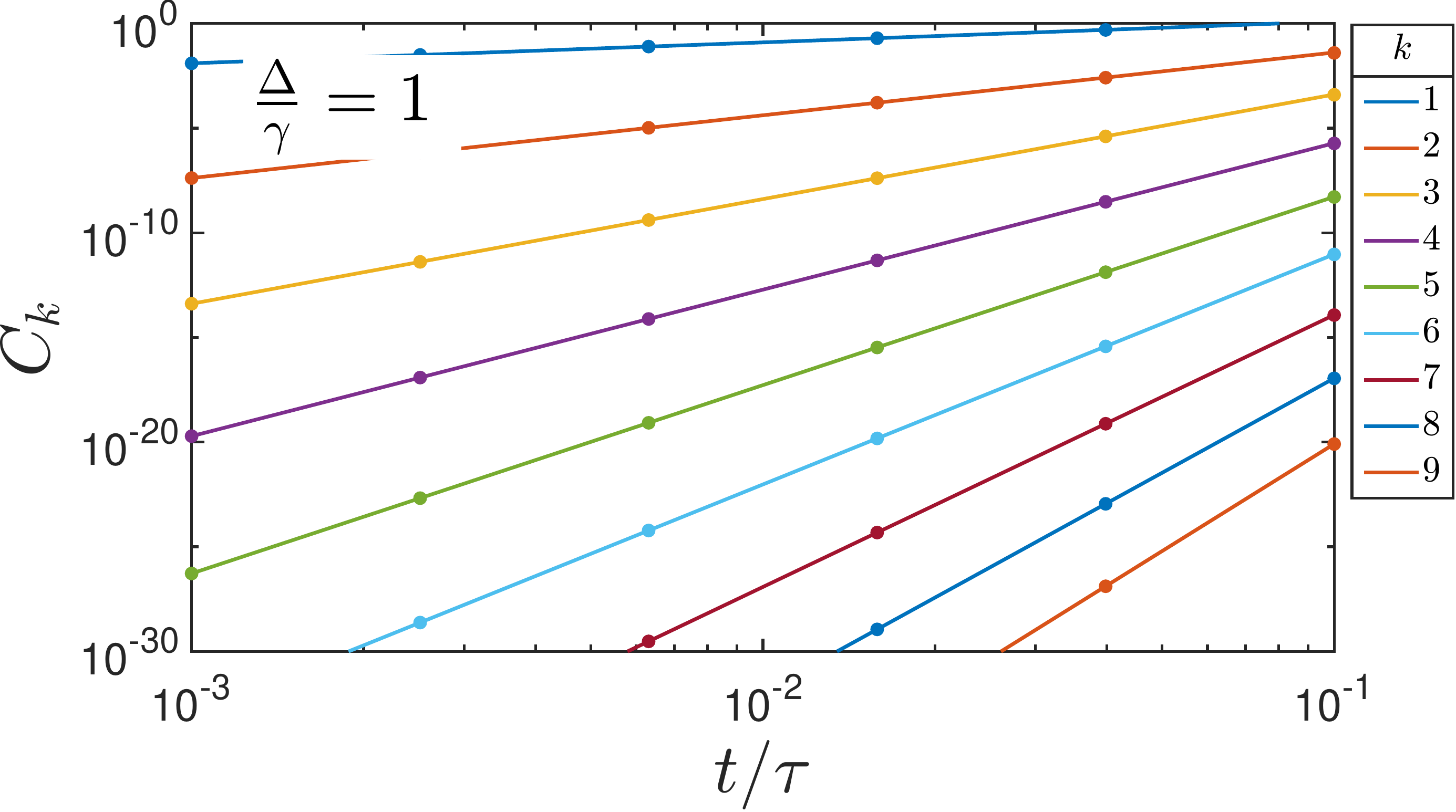}
\caption{The Lieb-Robinson early-time correlation function for the nine-qubit line demonstrates the power-law dependence on time. The qubit index is $k$ and the dots represent  numerical evaluation of Eq. (\ref{eq:LRcorrExact}) for the Hamiltonian in Eq. (\ref{eq:HIsing}) with $\Delta/\gamma=1$. The lines represent the analytic result of Eq. (\ref{eq:LR_AnalyticChain}). 
}
\label{fig:LRloglogDG1}
\end{figure}

\begin{figure}[tb]
\centering
\includegraphics[width=8.6cm]{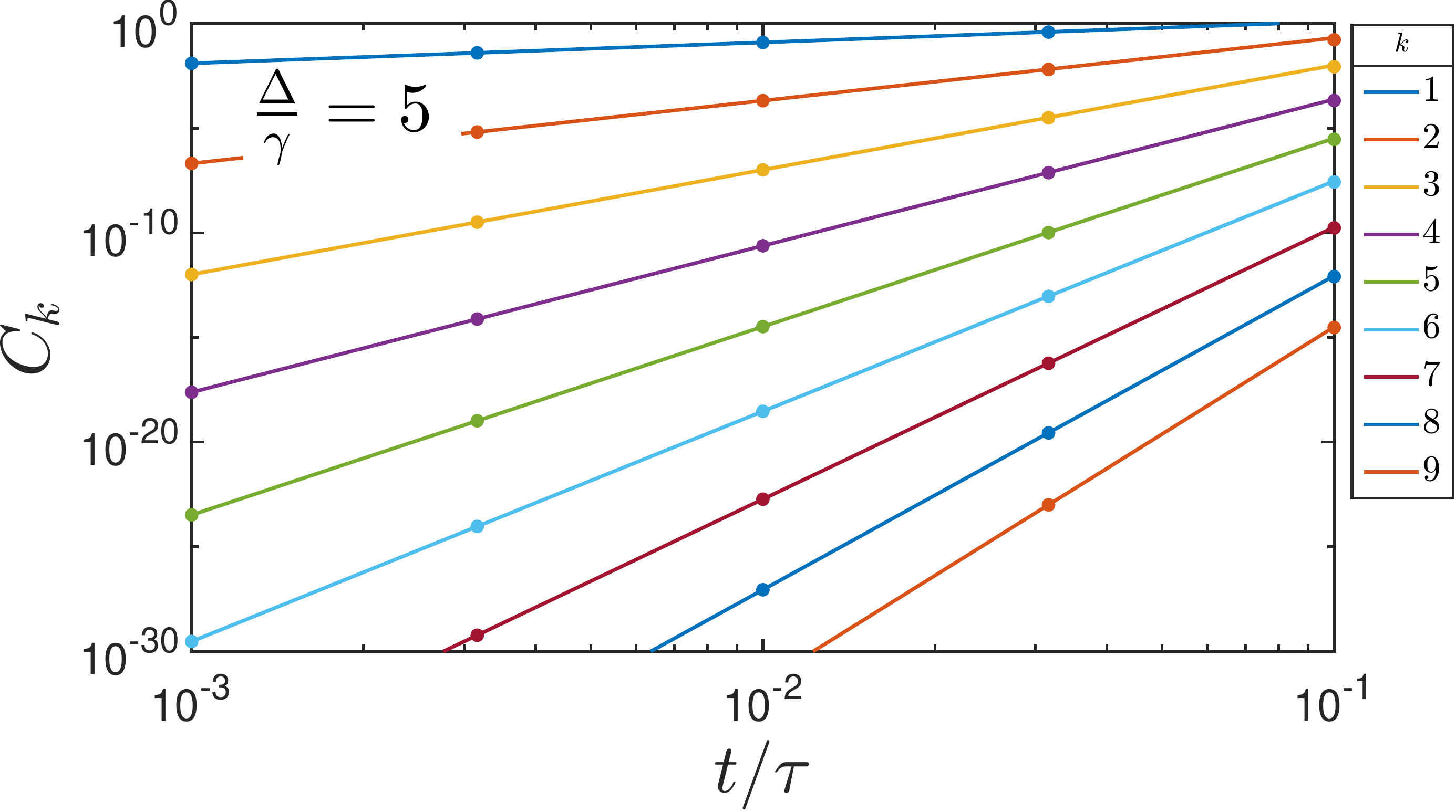}
\caption{The Lieb-Robinson early-time correlation function for the nine-qubit line with qubit index $k$. The dots represent the direct numerical evaluation of Eq. (\ref{eq:LRcorrExact})  with $\Delta/\gamma=5$ and the lines represent the analytic result of Eq. (\ref{eq:LR_AnalyticChain}). The stronger coupling between neighboring qubits yields a faster propagation of correlations down the chain.
}
\label{fig:LRloglogDG5}
\end{figure}

The early time behavior of the Lieb-Robinson correlation function has a power-law dependence, as is evident in Fig. \ref{fig:LRCorrZoomIn} and in the log-log plot shown in Fig. \ref{fig:LRloglogDG1}. The dots in the figure represent the numerical solution of equation (\ref{eq:LRcorrExact}) for $\Delta/\gamma=1$. 
The lines are the results of the analytical expression derived below. 
Figure \ref{fig:LRloglogDG5} similarly shows the numerical result for the case when $\Delta/\gamma=5$.  The larger interaction strengths makes the correlations spread more rapidly down the line.

\subsection{Analytic solution for early-time behavior}
We adopt the following notation for an $n$-nested commutator.
\begin{equation}
    \left[(\hat{A})^n,\hat{B} \right] \equiv 
    \left[ \hat A, \left[\hat A, \ldots \left[ \hat A [\hat A, \hat B] \right] \right]\ldots \right]
    \label{eq:NestedCom}
\end{equation}

\noindent It will also be useful to define a notation for a nested commutator with a sequence of operators.
\begin{equation}
    \left[(\hat{\mathcal{O}}_1, \hat{\mathcal{O}}_2, \ldots, \hat{\mathcal{O}}_n),\hat{B} \right] \equiv 
    \left[ \hat{\mathcal{O}}_1, \left[ \hat{\mathcal{O}}_2,, \ldots \left[ \hat{\mathcal{O}}_n, \hat B] \right] \right]... \right]
    \label{eq:NestedListCom}
\end{equation}

Using the notation of Eq. (\ref{eq:NestedCom}), we can write the time dependence of any Heisenberg operator as
\begin{equation}
    \hat{A}(t) =  \hat{A}(0)+  \sum_{n = 1}^{\infty} \frac{1}{n!}\left(\frac{i t }{\hbar}\right)^n \left[(\hat{H})^n,\hat{A} \right].
    \label{eq:HeisenbergTime}
\end{equation}

\noindent Applying this to  $\hat{\sigma}_z^{(1)}(t)$  we have

\begin{eqnarray}
    \hat{C}_k(t) &=&  \left[ \left( \sum_{n = 1}^{\infty} \frac{1}{n!}\left(\frac{i t }{\hbar}\right)^n \left[(\hat{H})^n,\hat{\sigma}^{(1)}_z \right] \right),\hat{\sigma}^{(k)}_z  \right] \\
    &=& \sum_{n = 1}^{\infty} \frac{1}{n!}\left(\frac{i t }{\hbar}\right)^n \left[  \left[(\hat{H})^n,\hat{\sigma}^{(1)}_z \right] ,\hat{\sigma}^{(k)}_z  
     \right] \\
     &=&  \sum_{n = 1}^{\infty}  \hat C_k^{(n)} \left( \frac{t}{\tau} \right)^n
     \label{eq:LRoperatorSum}
\end{eqnarray}
The early time Lieb-Robinson operator for qubit $k$ is the lowest-order  $\hat C_k^{(n)}$ which is not zero. All other terms in Eq. (\ref{eq:LRoperatorSum}) will be vanishingly small for early enough times. 

For convenience, we define
\begin{equation}
    \hat G^{(n)}_k= \left[ \left[ (\hat H)^{(n)}, \hat\sigma_z^{(1)} \right],\hat\sigma_z^{(k)} \right],
    \label{eq:Gdef}
\end{equation}
\noindent so that,
\begin{equation}
    \hat C^{\,(n)}_k= \frac{1}{n!} \frac{\pi^n i^n}{\gamma^n} \hat G^{(n)}_k.
    \label{eq:C2G}
\end{equation}

To calculate the self-correlation (bit 1 with itself), we first evaluate
\begin{eqnarray}
    \left[(\hat H)^{(1)},\hat\sigma_z^{(1)} \right] &=& 
    (-\gamma)[\hat\sigma_x^{(1)},\hat\sigma_z^{(1)}] \nonumber\\
    && \;-\frac{\Delta}{2} [\hat\sigma_z^{(1)}\hat\sigma_z^{(2)}, \hat\sigma_z^{(1)} ]\nonumber\\
    &=& -(-2i)\gamma \hat\sigma_y^{(1)} 
\end{eqnarray}
\noindent so
\begin{eqnarray}
    \hat G^{(1)}_1 &=&\left[  \left[(\hat H)^{(1)},\hat\sigma_z^{(1)} \right], \hat\sigma_z^{(1)}\right] \nonumber\\
    &=& (-2i)^2 \gamma \hat\sigma_x^{(1)} 
\end{eqnarray}
\noindent and
\begin{eqnarray}
    \hat C^{\,(1)}_1 &=& -i4\pi \hat\sigma_x^{(1)}\nonumber\\
    C^{\,(1)}_1 &=& \|  \hat C^{\,(n)}_k \|= 4\pi. 
\end{eqnarray}
\noindent The early-time Lieb-Robinson self-correlation is therefore
\begin{equation}
    C_1(t) \approx   4 \pi \left( \frac{t}{\tau}\right) 
\end{equation}
\noindent where in this case the $n=1$ term is the first nonzero term which dominates at early times. The linear increase in $C_1(t)$ is evident in Figs. (\ref{fig:LRCorrZoomOut}), (\ref{fig:LRloglogDG1}), and (\ref{fig:LRloglogDG5}).

For the near-neighbor correlation $C_2(t)$,  the first non-zero term in Eq. (\ref{eq:Gdef}) occurs for $n=3$.
\begin{equation}
 \hat G^{(3)}_2= \left[ \left[ (\hat H)^{(3)}, \hat\sigma_z^{(1)} \right],\hat\sigma_z^{(2)} \right],   
 \label{eq:G3Line}
\end{equation}
\noindent The term $\left[ (\hat H)^{(3)}, \hat\sigma_z^{(1)} \right]$ contains many cross terms from the three iterated commutators with  $\hat H$ in Eq. (\ref{eq:HIsing}). But only one term survives the commutator with $\hat\sigma_z^{(2)}$ in Eq. (\ref{eq:G3Line}), with the result that
\begin{equation}
 \hat G^{(3)}_2=   (-2i)^4 (-\gamma)^2 \left( \frac{\Delta}{2} \right) 
                   \hat \sigma_x^{(1)} \hat\sigma_x^{(2)} 
 \label{eq:G3Line2}
\end{equation}
\noindent and the early-time result for the correlation function is
\begin{equation}
C_2(t)\approx    \frac{2^3}{3!}\pi^2 \left(\frac{\Delta}{\gamma} \right) 
              \left(\frac{t}{\tau} \right)^3.
 \label{eq:C3Line}
\end{equation}

For $C_3$, the $n=5$ term is the leading term. The structure of the {\em only} surviving term for $\hat G_3^{(5)}$ (Eq. (\ref{eq:Gdef})) in this case is illustrative. We  write it using the notation of Eq. (\ref{eq:NestedListCom}). 
\begin{equation}
    \hat G^{(5)}_3 \propto 
       \left[
        \left[\left( \hat\sigma_x^{(3)}, 
                \hat\sigma_z^{(3)} \hat\sigma_z^{(2)},
                \hat\sigma_x^{(2)}, 
                \hat\sigma_z^{(2)} \hat\sigma_z^{(1)},
                \hat\sigma_x^{(1)}
             \right),
             \hat\sigma_z^{(1)} 
             \right],
             \hat\sigma_z^{(3)}
             \right]
             \label{eq:G5line}
\end{equation}
\noindent The nested commutator has the structure $(X_3, \,\,\, Z_3 Z_2, X_2, \,\,\, Z_2 Z_1,  X_1)$  traversing the shortest path (directly) between qubit $3$ and qubit $1$. The alternation between terms requires 5 nestings of commutators with the the Hamiltonian. 
A nesting of this form  is necessary to avoid the whole expression becoming zero. The general feature is that to connect each pair of qubits along the path requires two nested commutators: a   $\hat\sigma_x^{(k)}$ term  followed by a $\hat\sigma_z^{(k)} \hat\sigma_z^{(k+1)}$ term. Therefore, only odd-order terms contribute.

Each commutator in Eq. (\ref{eq:G5line}) with $\hat\sigma_x$  generates a factor of $(-2i)\gamma$ and each commutator with the product $\hat\sigma_z \hat\sigma_z$ generates a factor of $(-2i)(-\Delta/2)$ so   
\begin{equation}
    \hat G^{(5)}_3 = (-2i)^6 \gamma^4 (-\Delta/2)^2  \hat\sigma_x^{(3)} 
             \hat\sigma_x^{(2)}\hat\sigma_x^{(1)}.
\end{equation}


\noindent The norm of the Pauli string is 1. Using (\ref{eq:C2G}), we then obtain
\begin{equation}
{C_3}(t) \approx   \frac{{{2^{4}}{\pi ^{5}}}}{{(5)!}}{\left( {\frac{\Delta }{\gamma }} \right)^{2}}{\left( {\frac{t}{\tau }} \right)^{5}}.
\end{equation}

In general, the lowest-order nonzero $\hat G^{(n)}_k$  is proportional to a Pauli string of $k$ $\hat\sigma_x$ terms, one for  each qubit between 1 and k, and $n=2k-1$.  
The  result for the early-time  Lieb-Robinson correlation function between the reference qubit 1 and qubit $k$  is given by:
\begin{equation}
{C_k}(t) \approx  \frac{{{2^{k + 1}}{\pi ^{2k - 1}}}}{{(2k - 1)!}}{\left( {\frac{\Delta }{\gamma }} \right)^{k - 1}}{\left( {\frac{t}{\tau }} \right)^{2k - 1}}.
\label{eq:LR_AnalyticChain}
\end{equation}

Correlations calculated with Eq. (\ref{eq:LR_AnalyticChain}) yield the solid lines in Figs. \ref{fig:LRloglogDG1} and \ref{fig:LRloglogDG5} and the dots show the values from  numerical solution of  Eq. (\ref{eq:LRcorrExact}). The numerical calculations include all orders in time, and thus are correct at all times,  not just early times, as illustrated in Fig. \ref{fig:LRCorrZoomOut}.  Equation  (\ref{eq:LR_AnalyticChain}) shows the leading order term, which is arbitrarily close to the exact result for early enough times. The excellent agreement between the two seen in Figs. \ref{fig:LRloglogDG1} and \ref{fig:LRloglogDG5} shows that the higher order terms are indeed negligible.

\subsection{Asymptotic behavior}
\begin{figure}[tb]
\centering
\includegraphics[width=8.6cm]{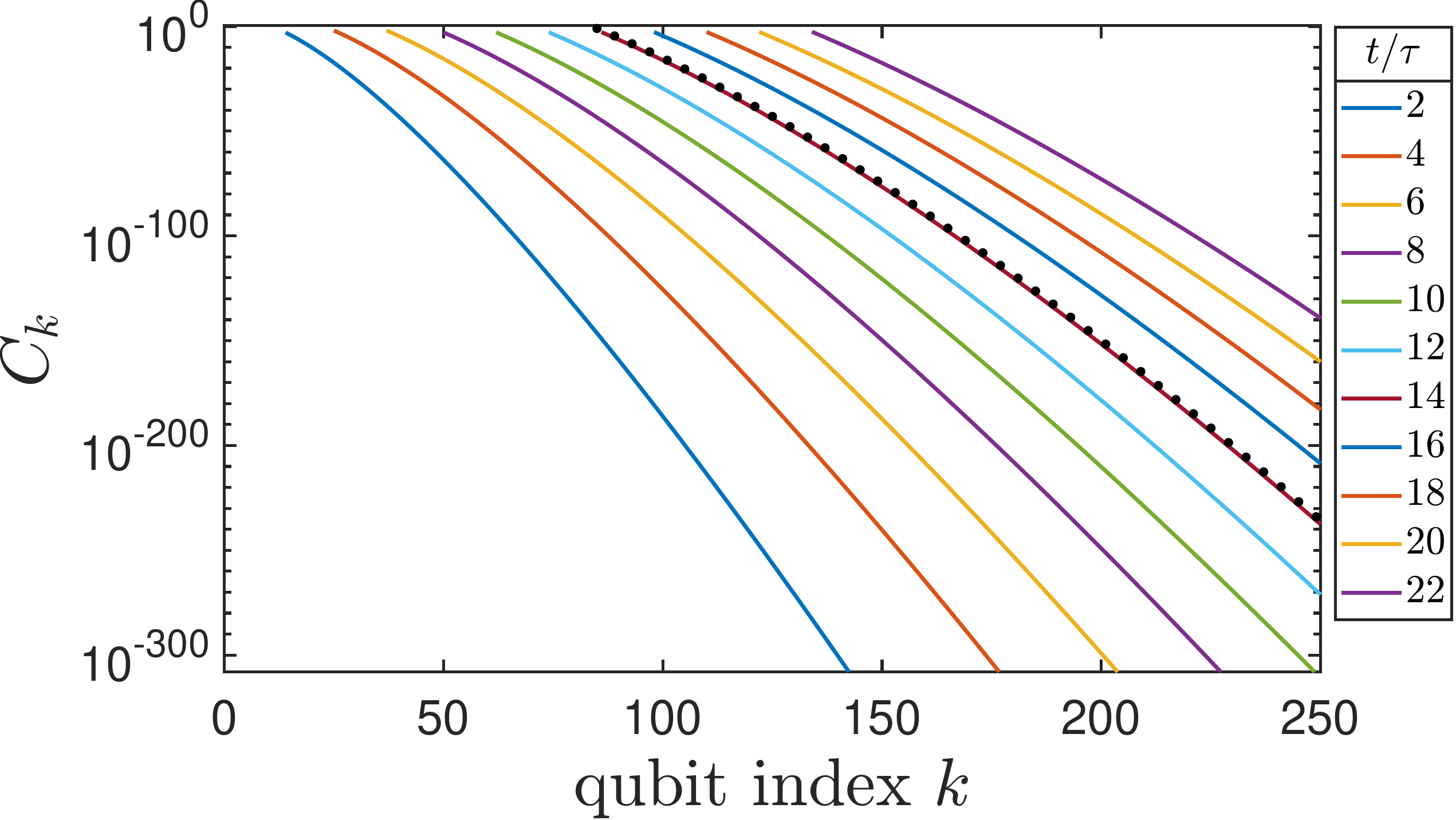}
\caption{Snapshots of the early-time Lieb-Robinson correlation at different times for a line of qubits as a function of qubit index $k$. The correlation front propagates down the line and, for the times shown, the shape of the front is attenuated faster than exponentially.  
The solid lines are the results of  Eq.(\ref{eq:LR_AnalyticChain}) with coupling $\Delta/\gamma=1$. The dotted line shows the correlation front at one time evaluated using Eq. (\ref{eq:LR1DlargeKlimit}).
}
\label{fig:CorrelationWavefrontEarly}
\end{figure}

Figures \ref{fig:LRCorrZoomOut} to \ref{fig:LRloglogDG5} show the Lieb Robinson correlation $C_k(t)$ for individual qubits with index $k$ in the linear array as a function of time. Figure \ref{fig:CorrelationWavefrontEarly} shows snapshots of the correlation function, computed from Eq. (\ref{eq:LR_AnalyticChain}), as a function of $k$ at particular times. This lets us see the ``correlation wave'' propagating from the reference bit down the line. Values  of $C_k$ larger than $10^{-2}$ are clipped and not shown here because our focus  is on early-time behavior for each qubit. At the times shown in the figure, the correlation front decays rapidly in space--faster than an exponential. 

We examine the asymptotic shape of the  correlation front propagating down the line quantitatively by taking the large $k$ limit of Eq. (\ref{eq:LR_AnalyticChain}).  The factorial can then be replaced by Stirlings approximation to obtain
\begin{equation}
    C_k(t)\approx  \sqrt{\frac{2}{\pi}} \sqrt{\frac{\gamma}{\Delta}} \frac{1}{\sqrt{k}}
    \left( \frac{v_{\text{\tiny LR}}}{(k-1/2)} \frac{t}{\tau}  \right)^{2k-1},
    \label{eq:LR1DlargeKlimit}
\end{equation}
where $v_{\text{\tiny LR}}$ is the Lieb-Robinson velocity given by
\begin{equation}
    v_{\text{\tiny LR}}=e\pi \sqrt{\frac{\Delta}{2\gamma}}.
    \label{eq:LR1Dvelocity}
\end{equation}
This velocity is expressed in terms of the natural dimensionless time $t/\tau$. In  dimensional form it would be $v_{\text{\tiny LR}}=(e/\sqrt{2}) \sqrt{\Delta \gamma}/\hbar$ qubits/s. 


\begin{figure}[tb]
\centering
\includegraphics[width=8.6cm]{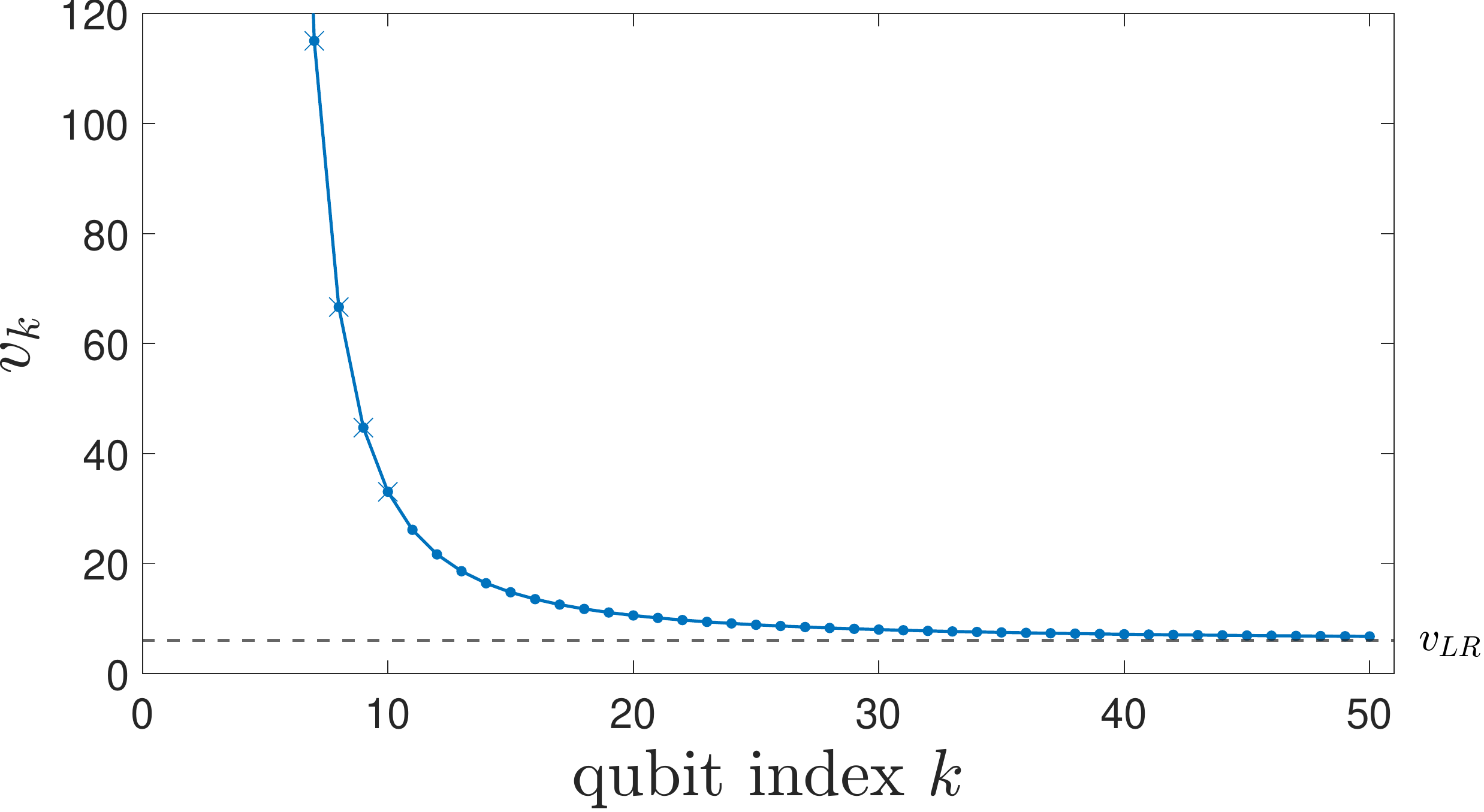}
\caption{Velocity saturation to the Lieb-Robinson velocity for the one-dimensional qubit line. The finite-difference velocity defined by Eq. (\ref{eq:VFiniteDifference}) is calculated from Eq. (\ref{eq:LR_AnalyticChain}) for for $\Delta/ \gamma= 1$ and $C_{\text{thresh}}=10^{-25}$. The values marked by an $\times$ are the results of numerical solution of Eq. (\ref{eq:LRcorrExact})
}
\label{fig:VelocitySaturation}
\end{figure}  

We can examine how the propagation of the correlation front converges to a constant speed. Let $C_{\text{thresh}}$  be a threshold value of the correlation function and let $t_{k}/\tau$ be the time at which qubit $k$ crosses that threshold of correlation, so $C_k(t_{k}/\tau)=C_{\text{thresh}}$. We define the backwards finite-difference velocity of the propagating front by
\begin{equation}
    v_k= \frac{1}{ t_k/\tau -t_{k-1}/\tau}.
    \label{eq:VFiniteDifference}
\end{equation}
Figure \ref{fig:VelocitySaturation} shows this velocity $v_k$ as a function of qubit index down the line. The velocity saturates to the value $v_{\text{\tiny LR}}$ given by Eq. (\ref{eq:LR1Dvelocity}). The line shows values calculated from Eq. (\ref{eq:LR_AnalyticChain}) for $\Delta/ \gamma= 1$ and $C_{\text{thresh}}=10^{-25}$. The saturated Lieb-Robinson velocity is then $e \pi /\sqrt{2} \approx 6.04$ in terms of the dimensionless time $t/\tau$. Figure \ref{fig:VelocitySaturation} also shows the values of $v_k$ calculated numerically directly from the defining  Eq. (\ref{eq:LRcorrExact}). 

\begin{figure}[tb]
\centering
\includegraphics[width=8.6cm]{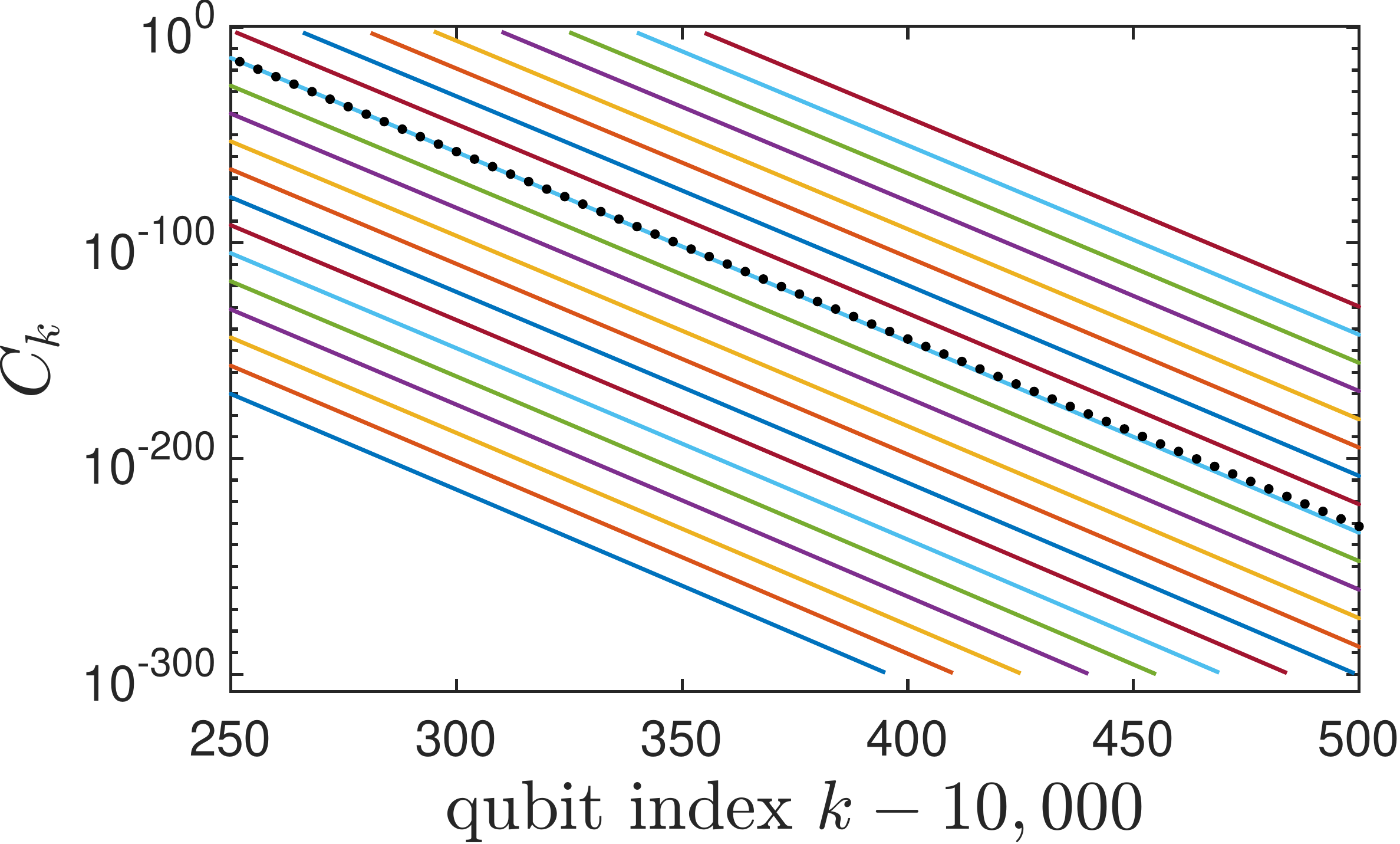}
\caption{Snapshots of the  Lieb-Robinson correlation as in Fig. \ref{fig:CorrelationWavefrontEarly}, but now for much later times and much farther down the qubit line.  Snapshots are shown for even values of $t/\tau$ between $1360$ and $1400$. Note that the qubit index $k$ is offset by $10^4$. By this point the correlation front, calculated from Eq. (\ref{eq:LR_AnalyticChain}), is well-approximated by the exponential dependence of Eq. (\ref{eq:LR1DexponentialLimit}) and propagates at the saturated Lieb-Robinson velocity given by Eq. (\ref{eq:LR1Dvelocity}). The dotted line shows the correlation front at one time evaluated in the limit of very large $k$ using Eq. (\ref{eq:LR1DexponentialLimit}).
}
\label{fig:CorrelationWavefrontLate}
\end{figure}  

Figure \ref{fig:CorrelationWavefrontLate} shows the correlation front much farther down the line (qubits with index more than $10,250$) and at considerably longer times ($t/\tau$ more than $1360$). By this point, the shape of the leading edge of the front is approximately exponential and moving at a steady speed. Taken together, Figs. \ref{fig:CorrelationWavefrontEarly} and \ref{fig:CorrelationWavefrontLate} also illustrate that ``early-time'' does not refer  to small values of $t/\tau$, but rather to the span of time before the quantum correlations between qubit $k$ and the reference qubit 1 become large.

In the region of the advancing correlation front for very large values of $k$, Eq. (\ref{eq:LR1DlargeKlimit}) can be written:
\begin{equation}
    C_k(t)\approx e \sqrt{\frac{2}{\pi}} \sqrt{\frac{\gamma}{\Delta}} \frac{1}{\sqrt{k}}
    e^{ -2 \left(k-v_{\text{\tiny LR}}\left(\frac{t}{\tau}\right) \right) }
    \label{eq:LR1DexponentialLimit}
\end{equation}
This matches the slope of the lines and the spacing in Fig. \ref{fig:CorrelationWavefrontLate}.

We note that the magnitude of early-time correlations become smaller as the front propagates down the chain because of the $k^{-1/2}$ dependence in Eq.(\ref{eq:LR1DexponentialLimit}). This also means the shape of the front differs slightly from a simple $e^{-2kx}$ in each snapshot shown in Fig. \ref{fig:CorrelationWavefrontLate}. The  attenuation of the Lieb-Robinson correlation function was also noted in Ref. \cite{LuitzPRR2020} for the  Heisenberg XXX model with short-range interactions.

\section{Correlation function for an arbitrary qubit array \label{sec:ArbArray}}
\begin{figure}[tb]
\centering
\includegraphics[width=5.6cm]{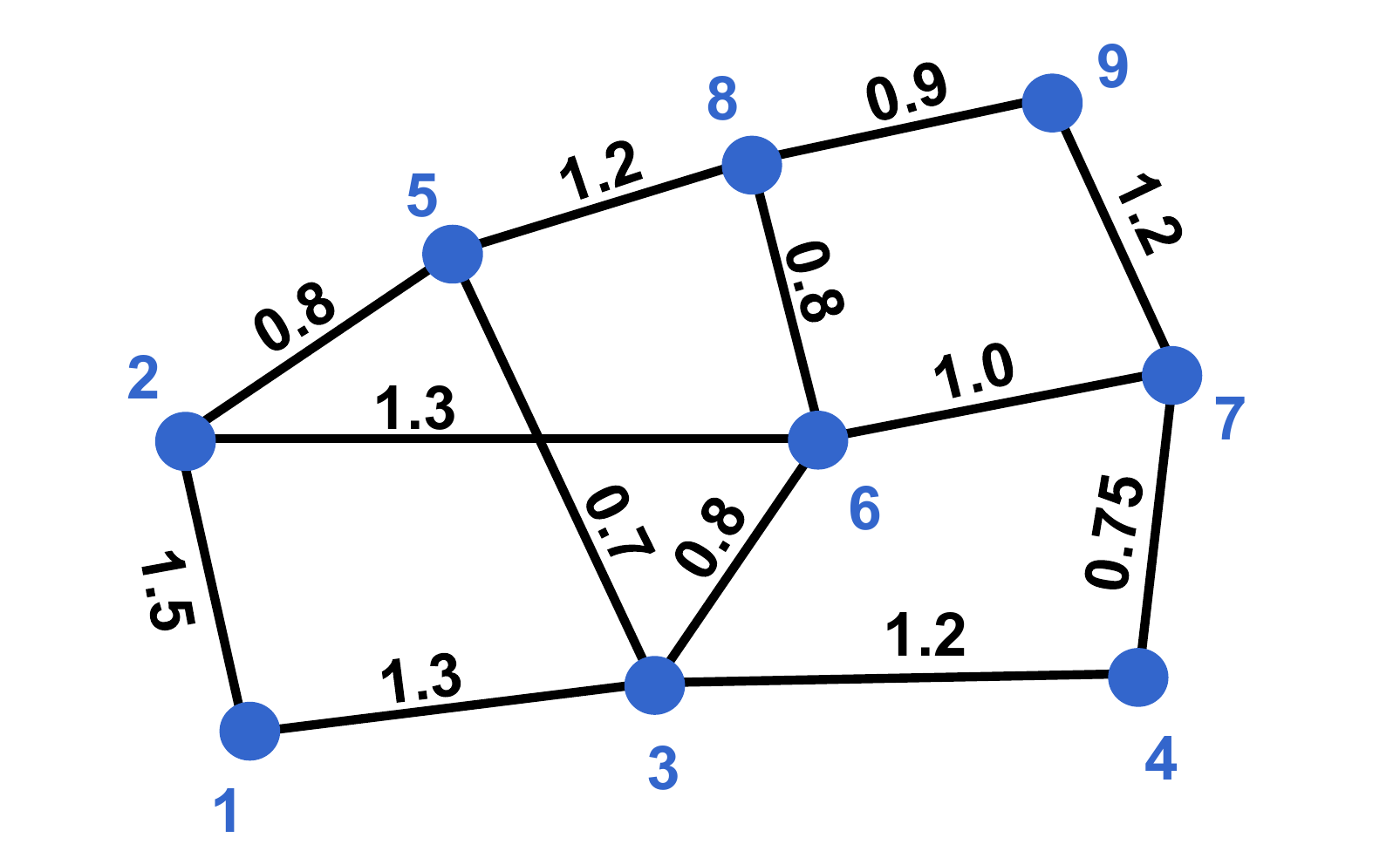}
\caption{An arbitrary network of qubits with different coupling parameters $\Delta$. 
}
\label{fig:ComplexNetworkDiagram}

\end{figure}

\begin{figure}[tb]
\centering
\includegraphics[width=8.6cm]{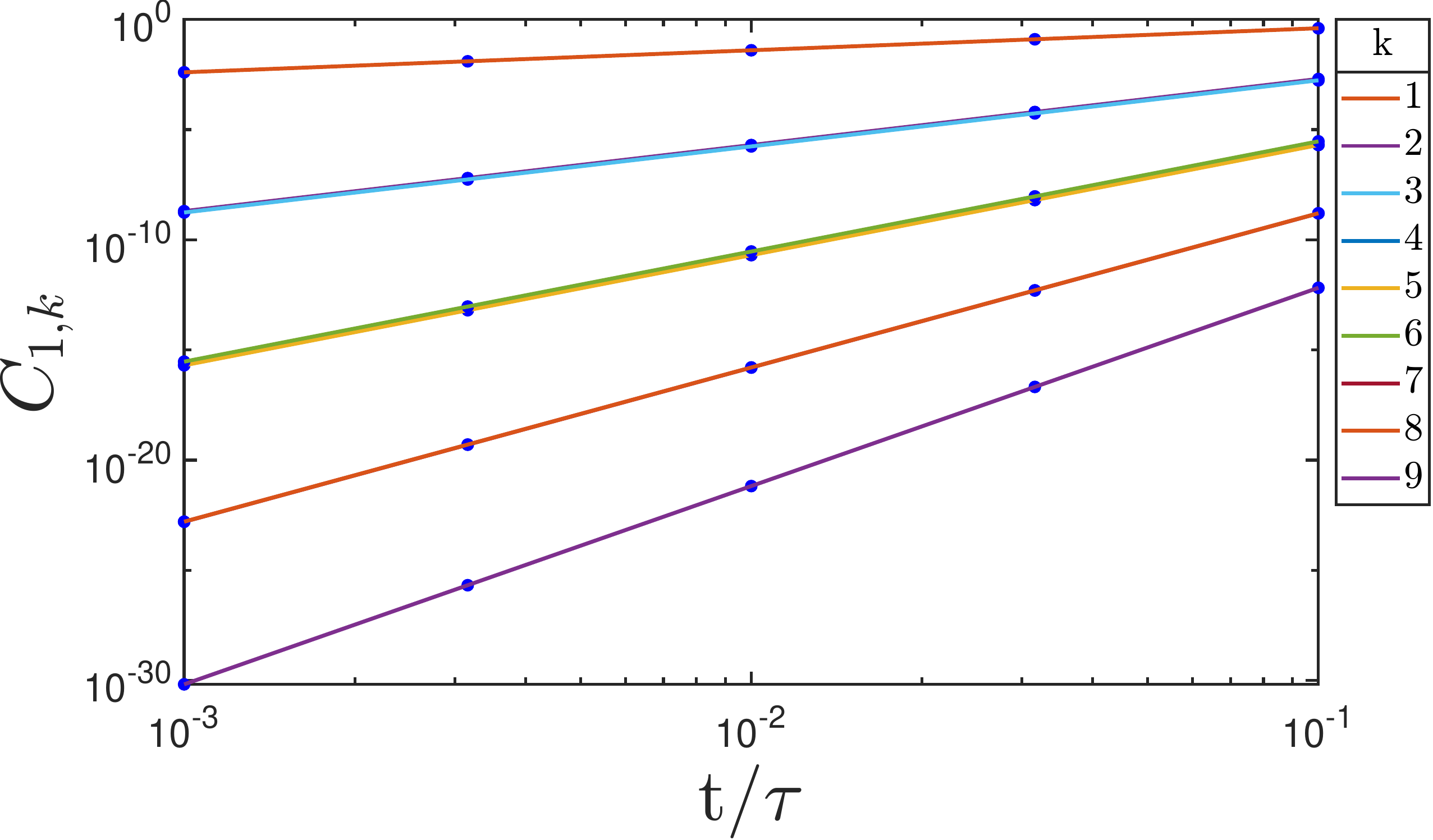}
\caption{The Lieb-Robinson correlation function for the qubit network shown in Fig. \ref{fig:ComplexNetworkDiagram}. $C_{1,k}$ is the correlation between qubit 1 and  qubit $k$. The dots are the result of   numerical evaluation of Eq. (\ref{eq:LRcorrExactArb}) and the lines represent the  general analytic result of Eq. (\ref{eq:GenLRcorrResult}). Results for some values of $k$ nearly overlap: $k=$2, 3; $k=$ 4, 5, 6; $k=$7, 8.  
}
\label{fig:ComplexNetworkCorr}
\end{figure}

The result in Eq. (\ref{eq:LR_AnalyticChain}) can be generalize for any network with arbitrary interaction strengths $\Delta_{j,k}$ between qubits $j$ and $k$. The Hamiltonian for the network is

\begin{equation}
    \hat{H} =  - \gamma \sum\limits_k^{{N_q}} {\hat \sigma _{_k}^x - } \frac{1}{2}\sum\limits_{j, k > j}^{{N_q}} 
        {{\Delta _{j,k}}\;  \hat \sigma_z^{(j)}} \,\hat \sigma_z^{(k)}.
        \label{eq:GenHamiltonian}
\end{equation}
Figure \ref{fig:ComplexNetworkDiagram} shows a  network of nine qubits with the strength of the coupling $\Delta_{j,k}$ between each indicated.  The dots in Figure \ref{fig:ComplexNetworkCorr} show the result of numerical calculation of the Lieb-Robinson correlation function between the first qubit and the $k^{th}$ qubit in the array using

\begin{equation}
    C_{1,k}(t) = \left\|  \left[ 
            e^{i\frac{\hat{H}}{\hbar} t} \hat{\sigma}_z^{(1)}
            e^{-i\frac{\hat{H}}{\hbar} t},
               \hat{\sigma}_z^{(k)}(0)\right] \right\|.
               \label{eq:LRcorrExactArb}
\end{equation}
The solid lines are the result of the analytic expression derived below. 


For the linear array studied in the previous section, there was one shortest path between qubit $1$ and qubit $k$, the direct path along the line,  that led to the  dominant term in $\hat G$ (Eq. \ref{eq:Gdef}). This was seen, for example, in the calculation for $G^{(5)}_3$ in Eq. (\ref{eq:G5line}), for which the path was simply $\{1, 2, 3\}$. We must now generalize this analysis to include more than one path through the network. A path is then a list of indices that begin at qubit $j$ and end at qubit $k$. The length of a path is the number of edges that connect successive qubits. 

For an arbitrary array of interacting qubits, there can be several paths with the same minimum length $L$ that contribute to the leading order early-time correlation function. We denote  each minimum-length path by $P_\alpha(j,k)$ where $\alpha$ is the index of the particular path. 
The index of the $m^{th}$ qubit along  path $P_\alpha(j,k)$ is denoted $q^\alpha_m$.   The lowest order nonzero operator 
\begin{equation}
    \hat G_{j,k}^{(2L+1)} = \left[ \left[ (\hat H)^{(2L+1)}, \hat\sigma_z^{(j)} \right],\hat\sigma_z^{(k)} \right],
\end{equation}
then becomes a weighted sum of Pauli strings of $\hat \sigma_x$ operators for each qubit traversed in the path. The  weights are given by the  products of the  couplings $\Delta_{q^{\alpha}_m,q^{\alpha}_{m+1}}$ along the path. We then obtain for the early-time Lieb-Robinson correlation function between qubits $j$ and $k$
\begin{multline}
C_{j,k}(t) \approx \frac{2^{L+2}\pi ^{2L+1}}{(2L+1)!} \left(\frac{t}{\tau}\right)^{2L+1}\\
\times\sqrt{\sum_{P_\alpha(j,k)} 
\left(\prod_{m = 1}^L \frac{\Delta_{q^{\alpha}_m,q^{\alpha}_{m+1}}}{\gamma}\right)^2}
\label{eq:GenLRcorrResult}
\end{multline}
\noindent where the sum is over all  paths with the same minimum length $L$. Equation (\ref{eq:GenLRcorrResult})  makes no assumption about the locality of the coupling $\Delta$ between qubits, which could extend far beyond near-neighbors.  Equation (\ref{eq:GenLRcorrResult})  reduces to Eq. (\ref{eq:LR_AnalyticChain}) when all the interaction strengths $\Delta$ are the same and  connect only near neighbors.   In that case, there is just the single minimum path between qubit $1$ and qubit $k$  which has length $L=k-1$. The results shown in Fig. \ref{fig:ComplexNetworkCorr} are in excellent agreement with the  numerical calculations.


\section{Correlation function for 2D and 3D regular lattices \label{sec:RegularLattices}}

\begin{figure}[bt]
\centering
\includegraphics[width=5.9cm]{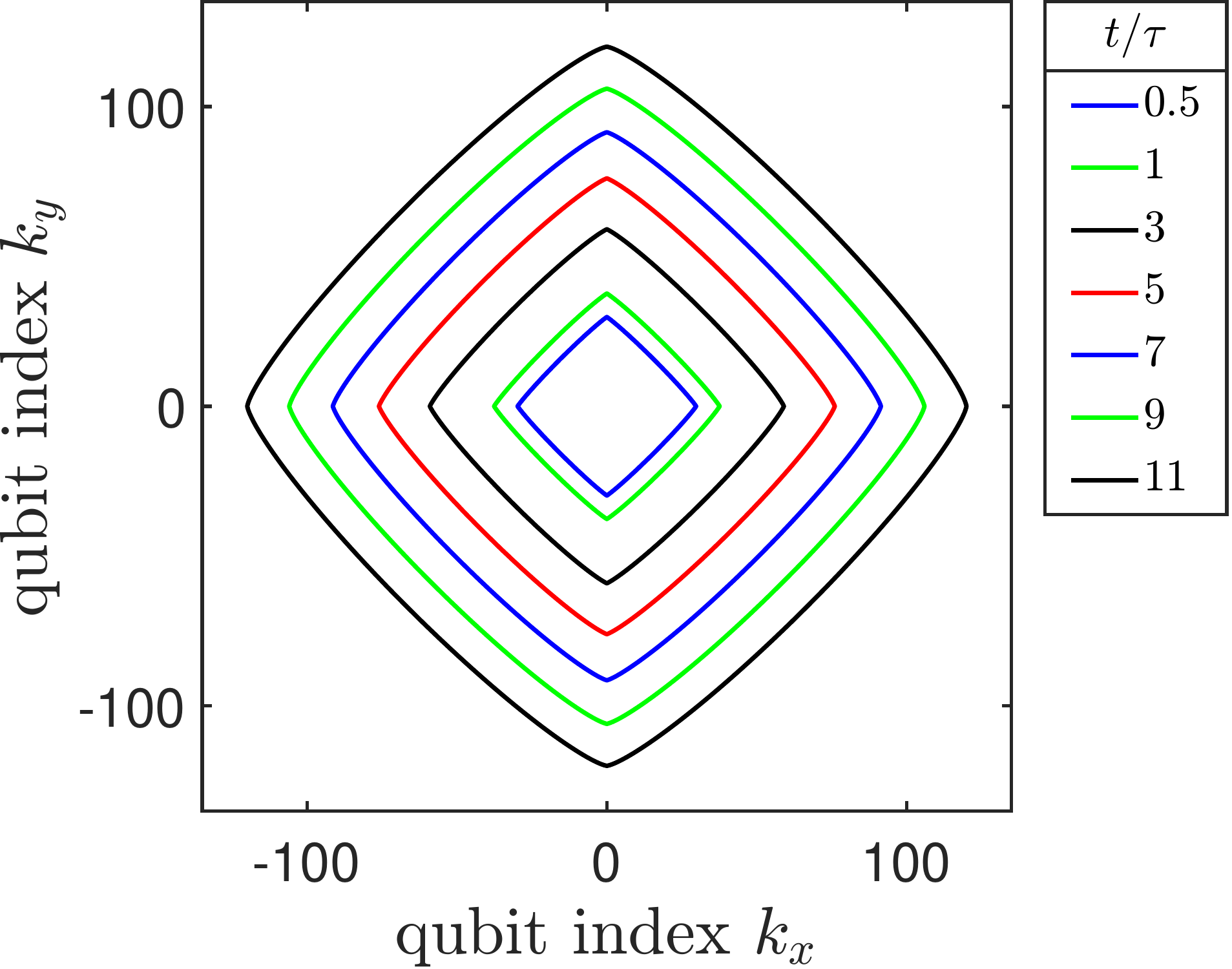}
\caption{Isocontours of the Lieb-Robinson correlation function for a square two-dimensional array of qubits with near-neighbor coupling. Contours are shown for correlation values of $10^{-60}$ at several values of time. 
}
\label{fig:2DLatticeFrobContours}
\end{figure}  

We consider a  two-dimensional square lattice of qubits with uniform near-neighbor coupling.  
\begin{multline}
\hat{H} =  - \gamma \mathlarger{\sum}\limits_{k_x,k_y} \hat{\sigma}^{(k_x,k_y)}_x\\
- \left(\frac{\Delta }{2}\right) 
\mathlarger{\sum}\limits_{\overset{\text{near neighbors}} {\overset{k_x,k_y}{  k'_x,k'_y}} } 
\hat{\sigma}^{(k_x,k_y)}_z \,\hat{\sigma}^{(k'_x,k'_y)}_z
\label{eq:HIsing2D}
\end{multline}
The qubit indices  $k_x$ and $k_y$ go from from $-N$ to $+N$. The reference qubit is at the origin $(k_x,k_y)=(0,0)$. Clearly there is no distinction between positive and negative directions, so the value of the Lieb-Robinson correlation function at a particular qubit  depends only on $(n,m)=(|k_x|,|k_y|)$. The minimum path length from the origin to $(n,m)$ is simply $m+n$, and the number of paths  with that length is $(n+m)!/(n!m!)$. 
Applying the general result of Eq. (\ref{eq:GenLRcorrResult}), we obtain for the correlation between the reference qubit at the origin and the $(n,m)$ qubit (in any quadrant): 
\begin{multline}
C_{n,m}(t) \approx \frac{2^{n+m+2}\;\pi^{2(n+m)+1}}{(2(n+m)+1)!} \sqrt{\frac{(n+m)!}{n!m!}} \\
 \times  \left(\frac{\Delta}{\gamma}\right)^{n+m} \left(\frac{t}{\tau}\right)^{2(n+m)+1}.
\label{eq:C2Dlattice}
\end{multline}


Figure \ref{fig:2DLatticeFrobContours} shown snapshots of isocontours of $C$ evaluated using Eq. (\ref{eq:C2Dlattice}) for several values of $t/\tau$. The Lieb-Robinson correlation front expands outward from the origin.  All the minimum-length paths are Manhattan paths, with the most direct route for expanding correlations along the coordinate axes. The correlation function along coordinate axes  reduces to the one-dimensional result  of Eq. (\ref{eq:LR_AnalyticChain}). (Note that the reference qubit is here indexed $(0,0)$ rather than 1.)

As with the one-dimensional case, the speed of the correlation front motion saturates to the Lieb-Robinson velocity, but for for the square 2D lattice the velocity depends on direction.  We  can express this in terms of the angle $\theta$  with the $x$-axis and find:
\begin{multline}
    v_{\text{\tiny LR}}(\theta) = e\pi \left(\sqrt{\frac{\Delta}{2\gamma}}\right)
     \left(\frac{1+\tan(\theta)}{\tan(\theta)^{\frac{\tan(\theta)}{\tan(\theta)+1}}}\right)^{\frac{1}{4}}\\
    \times \left(\frac{\sqrt{1 + \tan(\theta)^2}}{1 + \tan(\theta)}\right).
\end{multline}
This expression is valid for $0 \leq \theta \leq \pi/4$; other angles can be deduced by symmetry. 

\begin{figure}[bt]
\centering
\includegraphics[width=8.6cm]{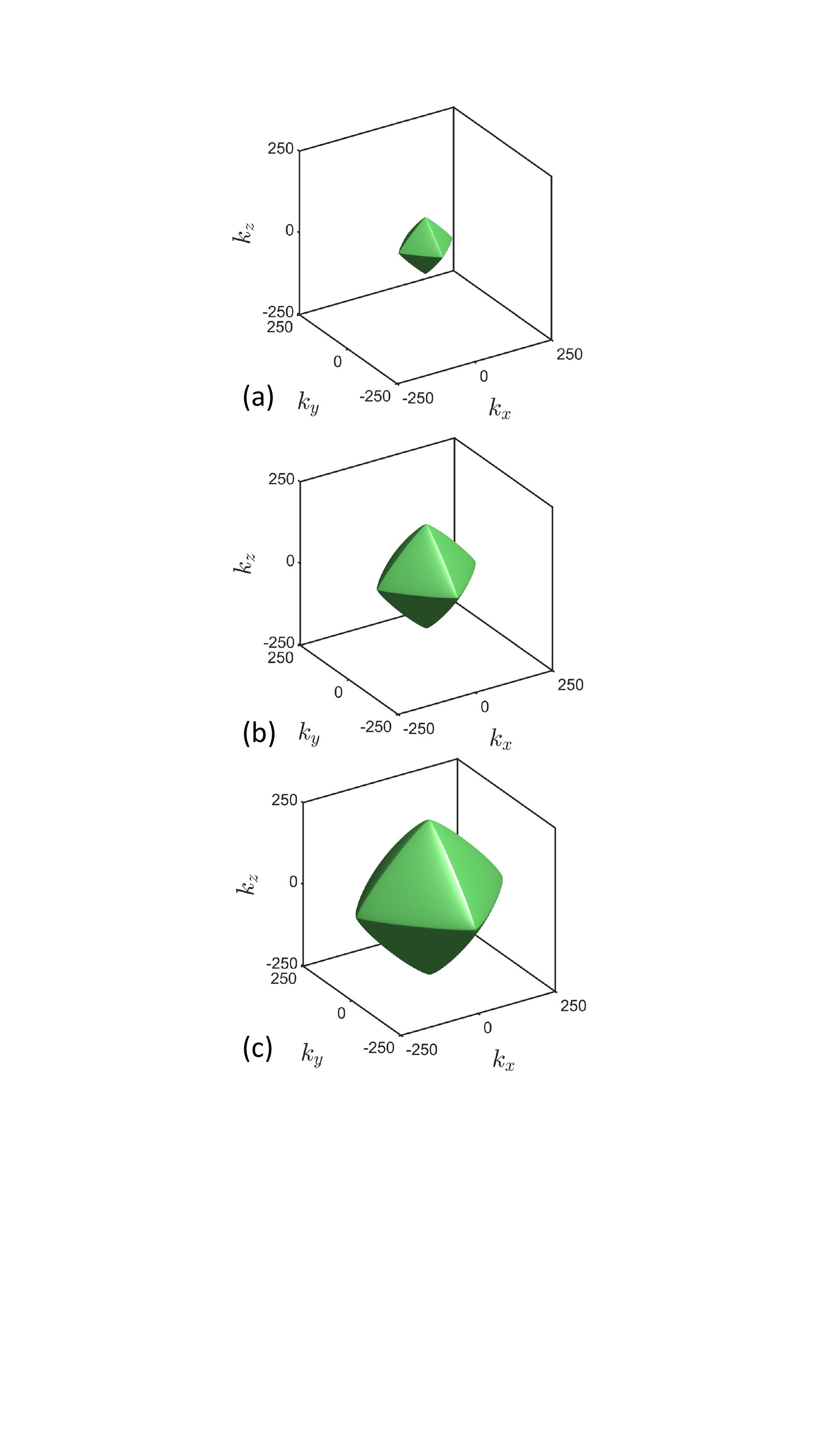}
\caption{Isosurfaces of the early-time Lieb-Robinson correlation function for a three-dimensional regular square lattice of qubits. The surfaces are shown for near-neighbor coupling $\Delta/\gamma=1$ and $C=1\times 10^{-170}$ at times (a)  $t/\tau=1$, (b) $t/\tau=5$, and (c) $t/\tau=11$. 
}
\label{fig:3DCorrFrob}
\end{figure}  

The Lieb-Robinson correlation for a three dimensional square lattice with uniform near-neighbor coupling can similarly obtained from the general result of Eq. (\ref{eq:GenLRcorrResult}). The correlation function between a qubit at the origin and one with coordinates $(n,m,p)=(|k_x|,|k_y|,|k_z|)$ (in any octant) can be written

\begin{multline}
C_{n,m,p}(t) \approx \frac{2^{n+m+p+2}\;\pi^{2(n+m+p)+1}}{(2(n+m+p)+1)!}\\ \times \sqrt{\frac{(n+m+p)!}{m!n!p!}}
\left(\frac{\Delta}{\gamma}\right)^{(n+m+p)}
\left(\frac{t}{\tau}\right)^{2(n+m+p)+1}.
\end{multline}

Figure \ref{fig:3DCorrFrob} shows snapshots of the isosurfaces of the Lieb-Robinson correlation function $C$ for three successive times. The influence of the Manhattan paths is clear in the rounded octahedral shape.

The direction-dependent Lieb-Robinson velocity for the 3D square lattice is expressed in terms of the polar angle $\phi$ and azimuthal angle $\theta$: 

\begin{multline}
    v_{\text{\tiny LR}}(\theta,\phi) = e \pi \left(\sqrt{\frac{\Delta}{2\gamma}}\right)
    \left( \frac{P(\theta,\phi)}{Q(\theta,\phi)} \right)^{\frac{1}{4}} R(\theta,\phi) 
\end{multline}
where
\begin{align}
    P(\theta,\phi)=1+\tan(\theta)+\sqrt{\frac{1+\tan(\theta)^2}{\tan(\phi)^2}}
\end{align}
\begin{multline}
    Q(\theta,\phi)=\tan(\theta)^
    {\left( \frac{\tan(\theta)}{1+\tan(\theta)+\sqrt{
     \frac{1+\tan(\theta)^2}{\tan(\phi)^2}} }\right)} \\
\times   \left( \sqrt{\frac{1+\tan(\theta)^2}{\tan(\phi)^2}} \right)
     ^{ \frac{\sqrt{\frac{1+\tan(\theta)^2}{\tan(\phi)^2}}}
      {1+\tan(\theta)+
     \sqrt{\frac{1+\tan(\theta)^2}{\tan(\phi)^2}}}}  
\end{multline}
and 
\begin{multline}
    R(\theta,\phi)=  \frac{\sqrt{1 +\tan^2(\theta) + \tan^2(\phi) +\tan^2(\theta)\tan^2(\phi) }}{\tan(\phi)+\tan(\phi)\tan(\theta)+\sqrt{1+\tan(\theta)^2}}
\end{multline}

\section{Discussion}
Calculating the value of the Lieb-Robinson correlation function itself, rather than a bound, for a specific Hamiltonian  illuminates some aspects of the way quantum information propagates. We are  able to see how the Lieb-Robinson velocity emerges as the saturation velocity of the correlation fronts propagation. A specific value for this velocity, and its angular dependence,  is also obtained for regular lattices.  We also see that the leading edge of the correlation front is super-exponential as the propagation starts, and acquires the exponential dependence in the bound of Eq. (\ref{eq:LRbound}) only later. 

Equation (\ref{eq:GenLRcorrResult}) is the main result of this work and applies to a large class of ZZ-coupled Hamiltionians given by Eq. (\ref{eq:GenHamiltonian}). We have specialized this result to address the case of regular square qubit arrays in 1D, 2D and 3D, with near-neighbor couplings.

There is some competition between the strength of coupling between qubits and the number of paths through intermediate qubits. But for early times the cost of adding path length is very high because of the $2L+1$ exponent in the time exponent of Eq. (\ref{eq:GenLRcorrResult}), and because of the factorial in the denominator. This latter factor also results in the decay of correlations in space as the front propagates outward. This is reflected the $k^{-1/2}$ in Eq. (\ref{eq:LR1DlargeKlimit}) for the Ising-coupled qubit line. 

\vfill

%

%

\end{document}